\documentstyle[12pt,fleqn,epsf,epsfig]{article}
\begin{document}
\title{ Origin of genetic information from minority control in a replicating system
with mutually catalytic molecules}
\author{
        Kunihiko Kaneko\\
        {\small \sl Department of Pure and Applied Sciences}\\
        {\small \sl University of Tokyo, Komaba, Meguro-ku, Tokyo 153, JAPAN}\\
       and\\
        Tetsuya Yomo\\
        {\small \sl Department of Biotechnology, Faculty of Engineering}\\
        {\small \sl Osaka University 2-1 Suita, Osaka 565, JAPAN}
\\}
\date{}
\maketitle
\vspace{-.3in}
\begin{abstract}
As the first step in an investigation of the origin of genetic information, 
we study how some species of molecules are preserved over cell generations and 
play an important role in
controlling the growth of a cell.  We consider a model consisting of protocells.  Each protocell
contains two mutually catalyzing molecule species ($X$ and $Y$), each of which has catalytically active 
and inactive types.  One of the species $Y$ is assumed to have
a slower synthesis speed.
Through divisions of the protocells, the system reaches and remains in
a state in which there are only a few active $Y$ and almost no inactive
$Y$ molecules in most protocells, through selection of very rare fluctuations.
In this state, the active $Y$ molecules are shown to control the
behavior of the protocell.  The minority molecule species act
as the information carrier, due to the relatively discrete nature of its population,
in comparison with the majority species which behaves
statistically in accordance with the law of large numbers.
The relevance of this minority controlled state to evolvability is discussed.
\end{abstract}


\pagebreak

\section{Introduction}

The origin of genetic information in a replicating system is an important theoretical 
topic that should be studied, not necessarily as a property of certain molecules,
but as a general property of replicating systems.
Consider a simple prototype cell that consists of mutually
catalyzing molecule species whose intra-cellular population growth results in
cell reproduction.  In this protocell, the molecules that carry the
genetic information are not initially specified, and to realize the growth in molecule numbers
alone, it
may not be necessary for specific molecules carrying such information to exist.

In actual cells, however, it is generally believed that information is encoded in
DNA, which controls the behavior of a cell.  
With regard to this point, though it is not necessary to take a
strong `geno-centric' standpoint,  it cannot be denied that there exists
a difference between DNA and protein molecules in the 
role of information carrier.
Still, even in actual cells, proteins and DNA both possess catalytic ability, and 
catalyze the production of each other, leading to cell replication.\footnote{
Note that through the reaction process from DNA to messenger RNA, and then to the
synthesis of proteins, a DNA molecule itself is not changed.  As a result, DNA works as a
catalyst for the synthesis of proteins.}  Then, why is
DNA regarded as the carrier of information?

To investigate this problem we need to clarify what it means for something to be an information carrier. 
For information carrying molecules, we identify the following two features as necessary. 

(1) If this molecule is removed or replaced by a mutant, there is a strong influence on the behavior of 
the cell.  We refer to this as the `control property'.

(2) Such molecules are preserved well over generations.  The number of such molecules exhibits
smaller fluctuations than that of other molecules, and their chemical structure (such as 
polymer sequence) is preserved over a long time span, 
even under potential changes by fluctuations through the synthesis of these molecules.
We refer to this as the `preservation property'.

According to the present understanding of [Alberts et al. 1997], 
changes undergone by DNA molecules are believed to exercise stronger influences
on the behavior of cells than other chemicals.  
With a higher catalytic activity,
a DNA molecule has a stronger influence on the behavior of a cell.
Also, a DNA molecule is transferred to offspring cells 
relatively accurately, compared with other constitutes of the cell.
Hence a DNA molecule satisfies the properties (1) and (2).

In addition, a DNA molecule is stable, and 
the time scale for the change of DNA, e.g., its replication process as well as its
destruction process, is slower.  Because of this relatively slow replication, the number of DNA 
molecules is smaller than the number of protein molecules.   For one generation of cells,
single replication of each DNA molecule occurs typically, while other molecules
undergo more replications (and decompositions).

The question we address in the present paper  is as follows.
Consider a protocell with mutually catalyzing molecules.
Then, under what conditions, does one molecule species begin to carry information
in the sense of (1) and (2)?  We show, under rather general
conditions in our model of mutually catalyzing system, that 
a symmetry breaking between
the two kinds of molecules takes place, and through replication and selection,
one kind of molecule comes to satisfy
the conditions (1) and (2).

Without assuming the detailed biochemical properties of DNA, we seek a
general condition for the differentiation of the roles of molecules  in a cell and 
study the origin of the controlling behavior of some molecules.
Assuming only a difference in the synthesis speeds of the two kinds of molecules,
we show that the species that eventually possesses a smaller population satisfies (1) and (2) and
acts as an information carrier.
With this approach, we discuss the origin of information from a kinetic viewpoint.
Note that we consider this information problem at a minimal level, i.e., as the origin of 1-bit
information in a replicating cell system.

In the present paper we consider a very simple protocell system (see Fig.1), 
consisting of two species of replicating molecules that catalyze each other.
Each species has active and inactive molecule types, with only
the active types of one species catalyzing the replication of the
the other species for the replication.
The rate of replication is different for the two species.
We consider the behavior of a system with
such mutually catalyzing molecules with different replication
speeds, as a first step in answering the question posed above.
We show that the molecule species with slower replication 
speed comes to possess the properties (1) and (2),
and that it therefore comes to represent the information carrier for cell replication.
Finally, we discuss why a system with a separation of 
roles between  information carrying and metabolism has a higher evolvability
(i.e., ability to evolve),  in reference to a recent experiment on an artificial replication system.

\section{Toy Model}

To study the general features of a system with mutually catalyzing molecules,
we consider the following minimal model.  First, we envision 
a (proto)cell containing molecules.  With a supply of
chemicals available to the cell, these molecules replicate through catalytic
reactions, so that their numbers within a cell increase.  When the total number
of molecules exceeds a given threshold, the cell divides into two,
with each daughter cell inheriting half of the molecules of the mother, chosen randomly.
Regarding the chemical species and the reaction, we make the
following simplifying assumptions:

(i) There are two species of molecules, X and Y, which are mutually catalyzing.

(ii) For each species, there are active (A) and inactive (I) types.  There are thus four types,
$X^A$, $X^I$, $Y^A$, and $Y^I$.
The active type has the ability to catalyze the replication of both types of
the other species of molecules. The catalytic reactions for
replication are assumed to take the form\footnotemark

\begin{math}
X^J + Y^A \rightarrow 2 X^J +Y^A\end{math} (for $J=A$ or $I$)

and 

\begin{math}
Y^J + X^A \rightarrow 2 Y^J +X^A\end{math} (for $J=A$ or $I$).

\footnotetext{More precisely, there is a supply of precursor molecules for the
synthesis of $X$ and $Y$, and the replication occurs with catalytic influence of 
either $X^A$ or $Y^A$.}

(iii) The rates of synthesis (or catalytic activity) of
the molecules $X$ and $Y$ differ.  We stipulate that the rate of the above replication process 
for $Y$,
$\gamma_y$, is much smaller than that for $X$, $\gamma_x$.
This difference in the rates may also be caused by a difference in
catalytic activities between the two molecule species.

(iv) It is natural to assume that the active molecule type
is rather rare.  With this in mind, we assume that there are $F$ types
of inactive molecules per active type.  For most simulations,
we consider the case in which there
is only one type of active molecules for each species.

(v) In the replication process, there may occur
structural changes that alter the activity of molecules. Therefore the
type (active or inactive) of a daughter molecule can differ from that of the
mother.  The rate of such structural change is given by $\mu$, which is not necessarily small,
due to thermodynamic fluctuations. 
This change can consist of the alternation of a sequence in a polymer or other conformational
change, and may be regarded as replication `error'.
Note that the probability for the loss of activity is $F$ times greater than for its
gain, since
there are $F$ times more types of inactive molecules than active molecules.
Hence, there are processes described by

\begin{math}
X^I \rightarrow X^A;\end{math}and \begin{math}Y^I \rightarrow Y^A\end{math} (with rate $\mu$)

\begin{math}
X^A \rightarrow X^I;\end{math}and \begin{math}Y^A \rightarrow Y^I\end{math}(with rate $\mu F$),

resulting from structural change.

(vi)When the total number of molecules in a protocell exceeds
a given value $2N$, it divides into
two, and the chemicals therein are distributed  into the two daughter cells randomly,
with $N$ molecules going to each.
Subsequently, the total number of molecules in each daughter cell increases from $N$ to
$2N$, at which point these divide.

(vii) To include competition, we assume that there is a constant total
number 
$M_{tot}$ of protocells, so that one protocell, randomly chosen,
is removed whenever a (different) protocell divides into two.

With the above described process, we have basically four sets of parameters:
the ratio of synthesis rates $\gamma_y/\gamma_x$,
the error rate $\mu$, the fraction of active molecules $1/F$, 
and the number of molecules $N$.  (The number $M_{tot}$ is not
important, as long as it is not too small).

We carried out simulation of this model, according to the following procedure.
First, a pair of molecules is chosen randomly. 
If these molecules are of different species, then if the
$X$ molecule is active, a new $Y$ molecule is produced with the probability
 $\gamma_y$, and if the $Y$ molecule is active, a new $X$ 
molecule is produced with the probability $\gamma_x$.
Such replications occur with the error rates given above.
All the simulations were thus carried out
stochastically, in this manner.

We consider a stochastic model rather than the
corresponding rate equation, which is valid for large $N$, since we are interested in
the case with relatively small $N$.  This follows from the fact that in a
cell, often the number of molecules of a given species is not large, and thus the
continuum limit implied in the rate equation approach is not necessarily justified
[Hess and Mikhailov 1994, 1995; Stange, Mikhailov, and Hess 1998].  
Furthermore, it has recently been found that the discrete nature of a molecule population
leads to qualitatively different
behavior than in the continuum case in a simple autocatalytic reaction network
[Togashi and Kaneko 2001].

\section{Result}

If $N$ is very large, the above described stochastic model can be replaced by a
continuous model given by the rate equation.
Then the growth dynamics of the number of molecules
$N_x^J$ and $N_y^J$ (for $J=A$ or $I$)
is described by the rate equations

\begin{equation}
dN_x^J/dt=\gamma_x N_x^J N_y^A;
dN_y^J/dt=\gamma_y N_x^A N_y^J.
\end{equation}
From these equations, under repeated divisions,
it is expected that the relations $\frac{N_x^A}{N_y^A}=\frac{\gamma_x}{\gamma_y}$,
$\frac{N_x^A}{N_x^I}= \frac{1}{F}$, and  $\frac{N_y^A}{N_y^I} = \frac{1}{F}$ are eventually satisfied.  
Indeed, even with our stochastic simulation, 
this number distribution is approached as $N$ is increased.

However, when $N$ is small, and with the selection process, 
there appears a significant deviation from the above distribution.
In Fig.2, we have plotted the average numbers $\langle N_x^A \rangle$, $\langle N_x^I \rangle$,
$\langle N_y^A \rangle$, and $\langle N_y^I \rangle$.  Here, $\langle ... \rangle$ represents the 
average over time of the number of the molecules of the individual species,
existing in a cell just prior to the division, when the total
number of molecules is $2N$, averaged over all observed divisions throughout the system.
(Accordingly, a cell removed without division does not contribute to the average).
As shown in the figure, there appears
a state satisfying $\langle N_y^A \rangle \approx 2 - 10$, $\langle N_y^I \rangle \approx 0$.
Since $F \gg 1$, such a state with $\frac{\langle N_y^A \rangle}{\langle N_y^I \rangle}>1$ 
is not expected from the rate equation (1).
Indeed, for the $X$- species, the number of inactive molecules is much larger than the number of
active ones.  Hence, we have found a novel state that can be realized
due to the
smallness of the number of molecules and the selection process.

In Fig.2, $\gamma_y/\gamma_x$ and $F$ are fixed to 0.01 and
64, respectively, while the dependence of \{$\langle N_x^A \rangle$,$\langle N_x^I \rangle$,$\langle N_y^A \rangle$,$\langle N_y^I \rangle$ \}
on these parameters is plotted in Figs. 3 and 4.
As shown in these figures, the above mentioned state with $\langle N_y^A \rangle \approx 2 - 10 $, $\langle N_y^I \rangle < 1$
is reached and sustained when $\gamma_y/\gamma_x$ is small and $F$ is sufficiently large.
In fact, for most dividing cells, $N_y^I$ is exactly 0, while there appear a few cells
with $N_y^I>1$ from time to time.
It should be noted that the state with almost no inactive Y
molecules appears in the case of larger $F$, i.e., in the case of
a larger possible variety of inactive molecules.  This suppression of 
$Y^I$ for large $F$ contrasts with the behavior found in the continuum limit (the rate equation).
In Fig.4, we have plotted $\frac{\langle N_y^A \rangle}{\langle N_y^I \rangle}$ as a
function of $F$.
Up to some value of $F$, the proportion of active $Y$ molecules decreases, 
in agreement with the naive expectation provided by Eq. (1),
but this proportion increases with further increase of $F$,  in the case that
$\gamma_y/\gamma_x$ is small ($\stackrel{<}{\sim}.02$) and $N$ is small. 

This behavior of the molecular populations 
can be understood from the viewpoint of selection: 
In a system with mutual catalysis, both
$X^A$ and $Y^A$ are necessary for the replication of
protocells to continue.  
The number of $Y$ molecules
is expected to be rather small, since their synthesis speed is much slower than that of
$X$ molecules.  Indeed, the fixed point distribution
given by the continuum limit equations possesses a rather small $N_y^A$.
In fact, when the total number of molecules is sufficiently small, the value
of $\langle N_y^A \rangle$ given by these equations is less than 1. 
 However, in  a system with mutual catalysis, both
$X^A$ and $Y^A$ must be present for replication of protocells to continue.
In particular, for the replication of $X$
molecules to continue, at least a single active $Y$ molecule is necessary.
Hence, if $N_y^A$ vanishes, only the replication of
inactive $Y$ molecules occurs.  For this reason, divisions producing descendants of this
cell cannot proceed indefinitely, because 
the number of $X^A$ molecules is cut in  half at each division.
Thus, a cell with $N_y^A<1$ cannot leave a continuing line of descendant
cells.  Also, for a cell with $N_y^A=1$, only one of its daughter cells can have an
active $Y$ molecule.  Hence a cell with $N_y^A=1$ has no 
potentiality to multiple through division, and for this reason, given 
the presence of cells with $N_y^A>1$ and selection,
the number of cells with $N_y^A=1$ should decrease with time.
We thus see that over a sufficiently long time, protocells with
$N_y^A>1$ are selected.

The total number of $Y$ molecules is limited to small values, due to their
slow synthesis speed.
This implies that a cell that suppresses the number of $Y^I$ molecules
to be as small as possible is preferable under selection,
so that there is a room for $Y^A$  molecules.
Hence, a state with almost no $Y^I$ molecules and a few $Y^A$ molecules,
once realized through fluctuations, is expected to be selected 
through competition for survival.

Of course, the fluctuations necessary to produce such a state decrease quite rapidly
as the total molecule number increases, and for sufficiently large numbers,
the continuum description of the rate equation is valid.
Clearly then, a state of the type described above is selected only when the total number of
molecules within a protocell is not too large. In fact, a state with 
very small $N_Y^I$ appears only if the total number $N$ is smaller than 
some threshold value depending on $F$ and $\gamma_y$.

To summarize our result, we have found that a
state with a few active $Y$ molecules and very small number of inactive $Y$ molecules  
is selected if the replication of $Y$ molecules is much slower than that of $X$, 
a large variety of inactive molecules exists, and
the total number of molecules is sufficiently small. 

{\sl Remark}:  In the model considered here, we have included a mechanism for the synthesis of
molecules, but not for their decomposition. To investigate the effect of the decomposition of
molecules, we have also 
studied a model including a process to
remove molecules randomly at some rate.  We found that the above stated conclusion is
not altered by the inclusion of this mechanism.

\section{Minority Controlled State}

In \S 3, we showed that in a mutually catalyzing replication system,
the selected state is one in which the number of inactive 
molecules of the slower replicating species, $Y$, is drastically suppressed.
In this section, we first show that the fluctuations of 
the number of active $Y$ molecules is smaller than those of active $X$ molecules in
this state. 
Next, we show that the molecule species $Y$
(the minority species) becomes dominant in
determining the growth speed of the (proto)cell system.  Then, considering a model
with several active molecule types, the control of chemical composition through
specificity symmetry breaking is demonstrated.

\subsection{Control of the growth speed}

First, we computed the time evolution of the number of active $X$ and $Y$ molecules,
to see if the selection process acts more strongly to control the
number of one or the other.  We computed $N_x^A$ and $N_y^A$  at every division
to obtain the histograms of cells with given numbers of active molecules.
(Here, the values of 
$N_x^A$ were coarse-grained into bins of size 10, chosen as
[0,10],[10,20],..., while all possible values of
$N_y^A$, 1,2,..., were computed separately.
The histograms for $N_x^A$ and $N_y^A$ were computed independently.

The histograms are plotted in Fig.6a.  We see that the distribution for $N_y^A$ has a sharp peak
near $N_y^A = 2$, while that for $N_x^A$ is much
wider.  Since the root mean square of a distribution increases
with the square root of the average for a standard random process, we have plotted 
the histograms by rescaling the ordinate by the expected average of
$\sqrt{N_x^A/N_y^A}$, which is approximately 10.  Even after this rescaling,
we find that the distribution of $N_x^A$ is much wider than that of $N_y^A$.
Hence, the fluctuations in the value of $N_y^A$ are much smaller than those of
$N_x^A$. We conclude that 
the selection process discriminates more strongly between different concentrations of 
active $Y$ molecules than between those of active $X$ molecules. 
Hence, it is expected that the growth speed of our protocell has
a stronger dependence on the number of active $Y$ molecules than the number of active $X$ molecules.

To confirm such a dependence, we have computed the 
dependence of the growth speed of a protocell 
on the numbers of molecules $X^A$ and  $Y^A$.  
Here, we computed $T_d$, the time required for division,
as a function of the number of active $X$ and $Y$ molecules at each division.
At every division we record $N_x^A$ and $N_y^A$ to determine
the time required for division $T_d$.  The division speed
for a given $N_x^A$ is computed as the division
time for this value of $N_x^A$ (with the bin size used for the histogram), 
averaged over all values of $N_y^A$, and similarly for a given
$N_y^A$. 
In Fig.6b, these average division times are
plotted as functions of $N_x^A$ and $N_y^A$.

As shown in the figure, the division time is a much more rapidly decreasing
function of $N_y^A$ than of $N_x^A$.  We see that even a slight change in the number of active $Y$ molecules has a strong influence on the
division time of the cell.  Of course, the growth
rate also depends on $N_x^A$, but this dependence is much weaker.
Hence, the growth speed is controlled mainly  by the active $Y$ molecules.

In addition, the fluctuations around this average division time
are smaller for fixed $N_y^A$.
To show this, we have computed the variance for $T_d^x(N_x^A)$ and $T_d^y(N_y^A)$.
Considering that the variance 
typically increases in proportion to the corresponding average,
we rescaled 
each variance by dividing by the corresponding average $\overline{T_d^x}(N_x^A)$ or $\overline{T_d^y}(N_y^A)$.
This scaled variance takes values around 0.55 for $T_d^x(N_x^A)$, and
around 0.25 for $T_d^y(N_y^A)$.
We thus conclude that the fluctuations of $T_d^y(N_y^A)$ for
fixed $N_y^A$ are smaller.
This implies that if $N_y^A$ is fixed, fluctuations of the division speed 
due to changes in $N_x^A$ are much smaller than the other way around.  
In other words,
the growth speed is controlled mainly by $N_y^A$.

\subsection{Preservation of the minority molecule}

As another demonstration of control, we study a model in which there is
more specific catalysis of
molecule synthesis.  Here, instead of single active molecule types 
for $X$ and $Y$, we consider a system with $k$ types of active $X$ and $Y$ molecules, $X^A(i)$ and $Y^A(i)$ ($i=1,2,\cdots k$).
In this model, each active molecule type catalyzes the synthesis
of only a few types ($m<k$) of the other species of molecules.
Graphically representing the ability for such catalysis using arrows as
$i_x \rightarrow j_y$ for $X \rightarrow Y$ and $i_y \rightarrow j_x$ for 
$Y \rightarrow X$, the network of arrows defining the catalyzing relations for
the entire system is chosen randomly, and is fixed throughout each simulation.  
An example of such a network (that which was used in the simulation discussed
below) is shown in Fig.7a.  Here we assume that both $X$ and $Y$ molecules 
have the same ``specificity" (i.e., the same value of $m$) and study how 
this symmetry is broken.

As discussed in \S 2, when $N$, $\gamma_y$ and 
$F$ satisfy the conditions necessary for realization of a state 
in which $N_y^I$ is sufficiently small, the surviving cell type contains
only a few active $Y$ molecules,
while the number of inactive ones vanishes or is very small.
Our simulations show that in the present model with several active molecules types,
only a single type of active $Y$ molecule remains after a sufficiently long time.  We call this ``remained type", $i_r$ ($1 \leq i_r \leq k$).
Contrastingly, at least $m$ types of $X^A$ species, that can be
catalyzed by the remaining $Y^A$ molecule species remain.
Accordingly, for a cell that survived after a sufficiently long time, 
a single type of $Y^A$ molecule catalyzes the synthesis of (at least) $m$ kinds of
$X$ molecule species, while the multiple types of $X$ molecules
catalyze this single type of $Y^A$ molecules.
Thus, the original symmetry regarding the catalytic specificity
is broken as a result of the difference between the synthesis speeds.

Due to autocatalytic reactions, there is a tendency for further 
increase of the molecules that are in the majority.  This
leads to competition for replication between molecule types
of the same species.  Since the total number of $Y$ molecules
is small, this competition leads to all-or-none behavior for the survival
of molecules. As a result, only a single type of species $Y$ remains,
while for species $X$, the numbers of molecules of different types are statistically distributed as guaranteed by the uniform replication
error rate.

The distribution of $X^A(i)$ species and the growth speed
depend on the identity of the remained type $i_r$ of $Y^A(i)$.
In Fig.7b, we display long-time number distributions of $X^A(i)$ molecules
reached from 6 different initial configurations, with a gray scale plot.  
The population distribution of $Y^A(i)$ molecules satisfies
$N(Y^A(i_r)) \approx (2 - 6)$, and $N(Y^A(j)) \approx 0$ for $j \neq i_r$.
The identity of the remaining type $i_r$ depends on the  initial conditions.
The number distribution of $X^A(i)$ and $X^I$ depends strongly on $i_r$,
as shown in Fig. 7b.  This strong dependence is
expected, since the $m$ types of $X$ molecules catalyzed by each active 
type of $Y$ molecule differ, as determined by the catalytic network (Fig.7a). 

Although $X$ and $Y$ molecules catalyze each other, a change in the type of
the remaining active $Y$ molecule has a much stronger influence on $X$ 
than a change in the types of the active $X$ molecules on $Y$,
since the number of $Y$ molecules is much smaller.
Consider, for example, a structural 
change of an active $Y$ molecule from type $i_r$ to $i_r'$ (for example,
the change in polymer sequence) that may occur 
during synthesis.  If such a change occurs and remains, there will be
a composition change from $N(Y(i_r)) \neq 0$ to $N(Y(i_r')) \neq 0$.
This change will alter the distribution of $X(i)$ drastically,
as suggested by Fig.7b.  By contrast, a structural change experienced by
$X$ molecules will have a much smaller influence on the distribution of
$Y(j)$. (This ignores the case in which many $X$ molecules change to 
a same type simultaneously by replication error,
resulting in a drastic change of the distribution of $X$. 
Such a situation, however, is very rare
in accordance with the law of large numbers).  In fact, there always
remain some fluctuations in
the distribution of $X$ molecules, while the distribution of $Y$ molecules
(i.e., identity of the remaining type $i_r$) is fixed over many generations, until 
a rare structural change leads to a different remaining type, which may
allow for a higher growth speed and the survival of the type containing it 
under selection.

With the results in \S 3 and \S 4, we can conclude
that the $Y$ molecules, i.e., the minority species, control the behavior of 
the system, and are preserved well over many generations.
We therefore call this state the minority-controlled (MC) state.

\subsection{Evolvability of the minority controlled state}

An important characteristic  of the MC state is evolvability.
Consider a variety of active molecules, with different catalytic activities.
Then the synthesis rates $\gamma_x$ and $\gamma_y$ depend on the activities of
the catalyzing molecules.  Thus, $\gamma_x$ can be written in terms of
the molecule's inherent growth rate, $g_x$, and the activity, $e_y(i)$, of
the corresponding catalyzing molecule $Y(i)$:

\begin{math}
\gamma_x =g_x \times e_y(i);
\gamma_y =g_y \times e_x(i).
\end{math}

\noindent
Since such a biochemical reaction is entirely facilitated by catalytic activity,
a change of $e_y$ or $e_x$, for example by the structural change of polymers,
 will be more important. Given the occurrence of such a change to molecules,
  those with greater catalytic activities will
be selected through competition evolution, leading to the selection of larger
$e_y$ and $e_x$.  As an example to demonstrate this point,
we have extended the model in \S 2 to include $k$ kinds of active molecules 
with different catalytic activities.  Then, molecules with greater catalytic
activities are selected through competition.

Here, the minority controlled state is relevant to realize evolvability. 
Since only a few molecules of the $Y$ species exist in the MC state,
a structural change to them strongly influences the catalytic activity of 
the protocell.  On the other hand, a change to $X$ molecules has a 
weaker influence, on the average, since the deviation of the {\sl average}
catalytic activity caused by such a change is smaller, as can be
deduced from the law of large numbers.
Hence the MC state is important for a protocell to realize
evolvability.

\section{Effect of Higher-order Catalysis}

In the first toy model considered in this paper, 
in order to realize the MC state, the difference between the time scales
of  the two kinds of molecules often must be rather large.
For example,  the ratio $\gamma_y/\gamma_x$ should typically be less than 
.05 when  the number of molecules is in the range $500 - 2000$.  
(If the number is larger, the rate should be much smaller.)
This difference in growth rates required to realize the MC state
is drastically reduced in a model that includes higher-order 
catalytic reaction processes in the replication  of molecules.

Consider a replication of molecules described by the following:

\begin{equation}
X+X'+Y \rightarrow 2X +X' +Y ; 
Y+Y'+X \rightarrow 2Y + Y' +X.
\end{equation}
In complex biochemical reaction networks, such higher-order catalytic
reactions often exist.
Indeed, proteins in a cell are catalyzed not solely by
nucleotides but with collaboration of proteins and nucleotides.
Nucleotides, similarly, are catalyzed not solely by proteins but
with collaboration of nucleotides and proteins.



In the continuum limit, the rate equation corresponding to the reaction (2) 
is given by $dN_x^j/dt =\gamma_x N_x^j N_x^A N_y^A$ and
$dN_y^j/dt =\gamma_x N_y^j N_y^A N_x^A$.
In this higher-order catalytic reaction, it is expected
that difference between the numbers of $X$ and $Y$ molecules is amplified.

Consider the equation $dx/dt=\gamma_x x^2y$ and
$dy/dt=\gamma_y y^2x$.  Then the relation

\begin{equation} 
x^{1/\gamma_x}/y^{1/\gamma_y}=const.=x_0^{1/\gamma_x}/y_0^{1/\gamma_y}
\end{equation}
holds, with the initial values $x_0$ and $y_0$ (where $x_0+y_0=C$).  
Then, consider the following division process:
if $x(t)+y(t)=2C$, then $x\rightarrow x/2$ and $y\rightarrow y/2$.
With the continued temporal evolution satisfying Eq.(3) and this
division process, $y$ approaches 0 if $\gamma_y < \gamma_x$.
(Recall that the curve $y(x)$ given by Eq.(3) is concave.)

Now, instead of the differential equations,
we again carried out stochastic simulations corresponding to
the above reaction process (2), using the same procedures 
(iii)-(vii) described in \S2.

The average values of active and inactive molecules obtained in these
simulations are
plotted in Fig.8 as functions of $\gamma_y$.
We see that if $\gamma_y / \gamma_x < .93$, the MC state is reached.
Note that here a difference in growth speeds as small as about 10\% 
is sufficient to realize the MC state.

In the present case, 
if $\gamma_y/\gamma_x$ is less than $\sim$.7, 
the system comes to exist in the state with $N_y^A=1,N_y^I=0$.
Since $N_y^A=1$, $X$ molecules
are synthesized, while $Y$ molecules are not.  Accordingly, after a division,
only the cell that inherits the $Y$ molecule keeps growing.
For this reason, the growth of the number of such protocells
is not possible, and hence a state with 
such a small value of $\gamma_y / \gamma_x$ is not expected to be reached through evolution.

The main conclusion of this section is that, when we consider
higher-order catalysis, the realization of the minority controlled state 
occurs for a wider range of values of  $\gamma_y / \gamma_x$.
In the above example,
a minority controlled state maintaining growth is realized
for $.7 < \gamma_y / \gamma_x < .93$, while the former inequality is always
satisfied as long as one considers a cell that continues to produce offspring.

\section{Discussion}

In this paper, we have shown that in a mutually catalyzing system,
molecules $Y$ with the slower synthesis speed tend to act as the
information carrier. Through the selection under reproduction, a state,
in which there is a very small number of inactive $Y$ molecules,
is selected.  This state is termed the ``minority controlled state".
Between the two molecule species, there appears separation of roles,
that with a larger number, and that with a greater catalytic activity.
The former provides a variety of chemicals and reaction paths, while
the latter holds ``information'', in the sense of
the two properties mentioned in the Introduction, `preservation' and `control'.
We now discuss these properties in more detail.  

[Preservation property]: A state that can be reached only through
very rare fluctuations is selected, and
it is preserved over many generations.\footnotemark    
In the theory presented here, the selected and preserved state is one
with $N_y^A \approx (2 - 10)$ and $N_y^I \approx 0$.
The realization of such a state is very rare
when we consider the rate equation obtained in the continuum limit.
For a model with several types of both molecule species,
the type of active $Y$ molecules with nonzero population 
remains fixed, in spite of the process of stochastic fluctuations.

\footnotetext{ Recall also  the
definition of the information by Shannon [Shannon and Weaver 1949],
according to which rarer events carry a greater amount of information. }

[Control property]: A change in the number of $Y$ molecules 
has a stronger influence on the growth rate of a cell than a change
in the number of $X$ molecules.
Also, a change in the catalytic activity of the $Y$ molecules has a strong influence on the growth of the cell.  The catalytic activity of the $Y$ 
molecules acts as a control parameter
of the system. 
For a model with several types of each molecule species,
$X$ (the majority species) has a smaller catalytic activity on the average,
and its catalysis is rather specific, only acting in the synthesis of
a single or a few types of molecules.
The minority species $Y$ has a greater catalytic ability and
acts to catalyze the synthesis of many kinds of molecules.  Hence a
change in $Y$ has a very strong influence.

With the information carrier defined in terms of
the preservation of rare states and control of the behavior of the system,
we have shown that molecule species with slower synthesis speed
acts as the information carrier.  In this way, the generation of information
is understood from a kinetic viewpoint.
Following our result, the separation of the roles of metabolism 
(``the chicken'') and information (``the egg") is explained as a general consequence of a cell system
with mutual catalysis and an appropriate difference between 
catalytic activities (leading to a difference in  synthesis speeds).

Finally, the following question remains: How does the difference
in the catalytic activity necessary to realize the MC state generally 
come to exist?
Of course, it is quite natural in a complex chemical system that 
there will be differences
in synthesis speeds or catalytic activities, and, in fact,
this is the case in the biochemistry of present-day organisms.
Still, it would be preferable to have a theory describing 
the spontaneous divergence of synthesis speeds without 
assuming a difference in advance, to provide a general model of
the possible `origin' of bio-information from any possible replication system.

To close the paper, we discuss (1) the evolutionary stability
and (2) evolutionary realizability of the MC state.

One important consequence of the existence of the MC state
is evolvability.  
Mutations introduced to the majority species tend to be cancelled
out on the average, in accordance with the law of large numbers.  Hence, the catalytic 
activity of the minority species ($Y$ in our model) is
not only sustained, but has a greater potentiality to increase 
through evolution.

Recently, there have been some experiments to construct minimal 
replicating systems
in vitro.  In particular, Matsuura et al. (2001) constructed a
replication system of molecules including DNA polymerase, synthesized by the corresponding gene.
Roughly speaking, the polymerase in the experiment corresponds to
$X$ in our model, while the polymerase gene corresponds to $Y$.  
In that experiment, instead of changing the synthesis speed $\gamma_y$ or $N$, 
the influence of the number of genes is investigated.

In the experiment,
it was found that replication is maintained even under deleterious mutations (that correspond to
structural changes from active to inactive molecules in our model),
only when the population of DNA polymerase genes is small and competition
of replicating systems is applied.
When the number of genes (corresponding to $Y$) is small, the
information containing in the DNA polymerase genes is preserved.
This is made possible by
the maintenance of rare fluctuations,
as found in our study.  

As discussed in \S 4, a change in catalytic activity can be included in 
the model by considering a system with several kinds of active molecules with different activities.
By considering a mutation from $X^A(i)$ to $X^A(j)$ (or $Y^A(i)$ to $Y^A(j)$),
accompanied by a change in the value of $e_x$ (or $e_y$),  one can examine 
the stability of the MC state with respect to mutation.  
If the initial difference between the catalytic abilities $e_x$ and $e_y$
(and other parameters) satisfies the conditions stated in \S 3,
the MC state is realized.  Then, we examined if such a state is destroyed 
by a change in the catalytic 
activities of molecules.  We found that this difference is
in fact maintained over many generations and that the MC state continues
to exist.  This behavior is due to the fact that a small mutation of $Y$
strongly influences the synthesis
of $X$, and a mutation resulting in a decrease of $e_y$ is not selected.
Hence the MC state possesses evolutionary stability.

The final remaining question we wish to address regards the
realizability of the MC state in the situation that
initially the two molecule species have almost the same catalytic activity.
One may expect that there would occur a divergence of the 
catalytic activities of two such molecule species, because once one species 
(say $Y$) has a larger catalytic activity, the number of $X$ molecules
will increase. This results in $Y$ becoming the minority species, 
which implies that its influences on the behavior of the cell will 
become stronger.  For this reason, the catalytic activity of
$Y$ increases faster than that of $X$, and thus the replication speed of $X$
becomes larger.  In this way, the difference between
replication speeds of $X$ and $Y$
might become further amplified.

While the above argument seems reasonable, it
does not hold for our model.
In simulations including such a structural change,
we have not observed such spontaneous symmetry breaking with regards
to the growth speeds of the two species, when these species initially
have (almost) equal catalytic activities.
The reason is as follows.  In our model,
the division of a cell is assumed to occur when the total
number of molecules becomes double the original number.  Now, in the
model, the collision of two molecules
is assumed to occur randomly. Hence the probability for a collision 
leading to synthesis of molecules should be
proportional to $\gamma_xN_xN_y+\gamma_yN_xN_y$,
if we assume a constant proportionality between the numbers of
active and inactive molecule.  Then, note that the quantity 
$N_xN_y\propto N_x(N-N_x)$
has a peak at $N_x=N_y$. It follows that the growth speed should be maximal
when the numbers of the two species are equal.
Hence there is a tendency toward a state in which there are
equal numbers of both species.  Of course, this argument is
rather rough, due to the assumption concerning the ratio of 
active to inactive molecule numbers, and the existence of a peak
at exactly $N_x=N_y$ may be slightly modified.  However,
the basic idea here is correct, and there is undoubtedly a tendency toward equal
numbers. For this reason, a state with a large difference is not reached 
spontaneously through some kind of symmetry breaking.

There are some possible scenarios within which the above described
tendency toward equal growth speeds may be ineffective.

1.  Higher-order catalysis:  As mentioned in \S 4, the 
imbalance necessary to realize a MC state is much smaller when higher-order
catalysis is considered.  Indeed, by introducing the mutation of 
catalytic activity to the model studied in \S 4, we have sometimes 
observed spontaneous symmetry breaking
between the parameters characterizing the two species.  The 
resulting state with a sufficient difference between the growth speeds
of two species, however, does not last very long,
since the necessary imbalance between $\gamma_x$ and $\gamma_y$ is so small 
that mutation can reverse the relative sizes of $\gamma_x$ and $\gamma_y$.

2.  Change in the collision condition:\footnotemark  In our model, collisions of molecules
occur randomly.  Hence if the number of $X$ molecules is larger,
most collisions occur between two $X$ molecules, and no reaction occurs.
However, if molecules are arranged spatially under different conditions
(e.g., consider the case in which $X$ molecules are on a membrane
and $Y$ molecules are in a contained medium), then the number of
reaction events between
$X$ and $Y$ molecules can be increased.  If we include this type of 
physical arrangement, which is rather natural when considering a cell, 
the tendency toward equal numbers no longer exists,
and the divergence of growth speeds in molecules should occur.

\footnotetext{Since the scenarios 2 and 3 assume another kind of
symmetry breaking between $X$ and $Y$ (albeit being different from the 
synthesis speeds), they cannot provide the final solution to the true spontaneous
symmetry breaking, although the assumptions may be 
biologically reasonable.}

3.  Condition for growth:  We have assumed that the division of the
protocell occurs when the total number of molecules doubles.
This assumption is useful as a minimal abstract model, but
it may be more natural to have a threshold that depends on
the number of molecules of one species (or, more generally, of some subset
of all species), rather than the total number.
For example, consider the case that division occurs when
the size of a membrane synthesized by biochemical reactions
is larger than some threshold.  This condition, for example, could be 
modeled by stipulating that division occurs when
the number of molecules of one species, say $X$, that composes the 
membrane, is larger than some threshold value.  
By imposing this type of division condition, the tendency toward
equal numbers of $X$ and $Y$ molecules could be avoided,
and the divergence of the
replication speeds of $X$ and $Y$ could take place.

4.  Network structure:  The catalytic network in a cell is generally 
quite complex, with many molecules participating in mutual catalysis
for replication. The evolution of replication systems with such catalytic 
networks have been studied since the proposal of the hypercycle by Eigen 
and Schuster [1979].  Dyson [1985], on the other hand, obtained a condition for
loose reproduction of protocells
with complex reaction networks consisting of active and inactive molecules. 
Origin of recursive replication from such loose reproduction is also discussed
[Segre, Ben-Eli, and Lancet 2000]. 

The differentiation of cells with a catalytic 
reaction network has also
been studied [Kaneko and Yomo 1997,1999; Furusawa and Kaneko 1998].
Here it has been found that
chemicals with low concentrations are often important in differentiation.
If the total number of molecules participating in the
reaction network is small, there should generally exist 
some species whose numbers of molecules are small, and
the discrete nature of these numbers plays a significant role.
For example, Togashi and Kaneko [2001] have found novel  
symmetry breaking that appears in a catalytic reaction system with 
a small total number of molecules.  Furthermore, 
a preliminary study of the reaction network version of the
model considered in this paper reveals spontaneous symmetry breaking 
that distinguishes a few controlling 
molecule species from a large number of non-controlling species,
without assuming a difference in synthesis speeds.

The symmetry breaking by the network structure is related 
with the evolution of specificity.
Although we have studied catalysis that has no specificity
(except for the model considered in \S 4.2), in reality 
one type of molecule can catalyze the synthesis of only a
limited number of molecule species.  Interestingly,
a preliminary study shows a symmetry breaking with regard to
the roles of molecules 
(with equal synthesis speeds) when higher-order catalytic reactions 
(as in \S 5) in random networks
with catalytic specificity are included.


\vspace{.1in}

{\bf Acknowledgment}

The authors would like to thank Shin'ichi Sasa and Takashi Ikegami
for useful discussions.
This research was supported by Grants-in-Aid for Scientific Research from
the Ministry of Education, Science and Culture of Japan (11CE2006,11837004).

\vspace{.1in}

{\bf References}

\begin{enumerate}

\item
B. Albert, D. Bray., J. Lewis, M. Raff, K. Roberts, and J.D. Watson {\sl The
Molecular Biology of the Cell} (1994)

\item
Freeman Dyson, {\sl Origins of Life}, Cambridge Univ. Press., 1985

\item
M. Eigen, P. Schuster,
{\sl The Hypercycle} (Springer, 1979).

\item
Furusawa C. \& Kaneko K.,
``Emergence of Rules in Cell Society: Differentiation, Hierarchy, and Stability"
Bull.Math.Biol.  60; 659-687 (1998)

\item
B. Hess and A. S. Mikhailov,
Science {\bf 264}, 223 (1994);

\item
B. Hess and A. S. Mikhailov,
J. Theor. Biol. {\bf 176}, 181 (1995).

\item
Kaneko K. \& Yomo T,
``Isologous Diversification: A Theory of Cell Differentiation ",
Bull.Math.Biol.  59, 139-196 (1997)

\item
Kaneko K. \& Yomo T,
``Isologous Diversification for Robust Development of
Cell Society ", J. Theor. Biol., 199 243-256 (1999)

\item
Matsuura T., Yomo T., Yamaguchi M, Shibuya N., Ko-Mitamura E.P., Shima Y., and
Urabe I. ``Importance of compartment formation for a self-encoding system",
preprint (2001)

\item
Segre D, Ben-Eli D, Lancet D.,
``Compositional genomes: prebiotic information transfer in mutually catalytic noncovalent assemblies'',
Proc Natl Acad Sci USA 97 (2000)4112-7

\item
C. Shannon and W. Weaver ``The Mathematical Theory of Communication",
Univ. of llinois Press, 1949

\item
P. Stange, A. S. Mikhailov and B. Hess,
J. Phys. Chem. B {\bf 102}, 6273 (1998)

\item
Y. Togashi and K. Kaneko,
`` Transitions Induced by the Discreteness of Molecules 
in a Small Autocatalytic System'' 
Phys. Rev. Lett. , 86 (2001) 2459

\end{enumerate}

\pagebreak

\begin{figure}
\noindent
\hspace{-.3in}
\epsfig{file=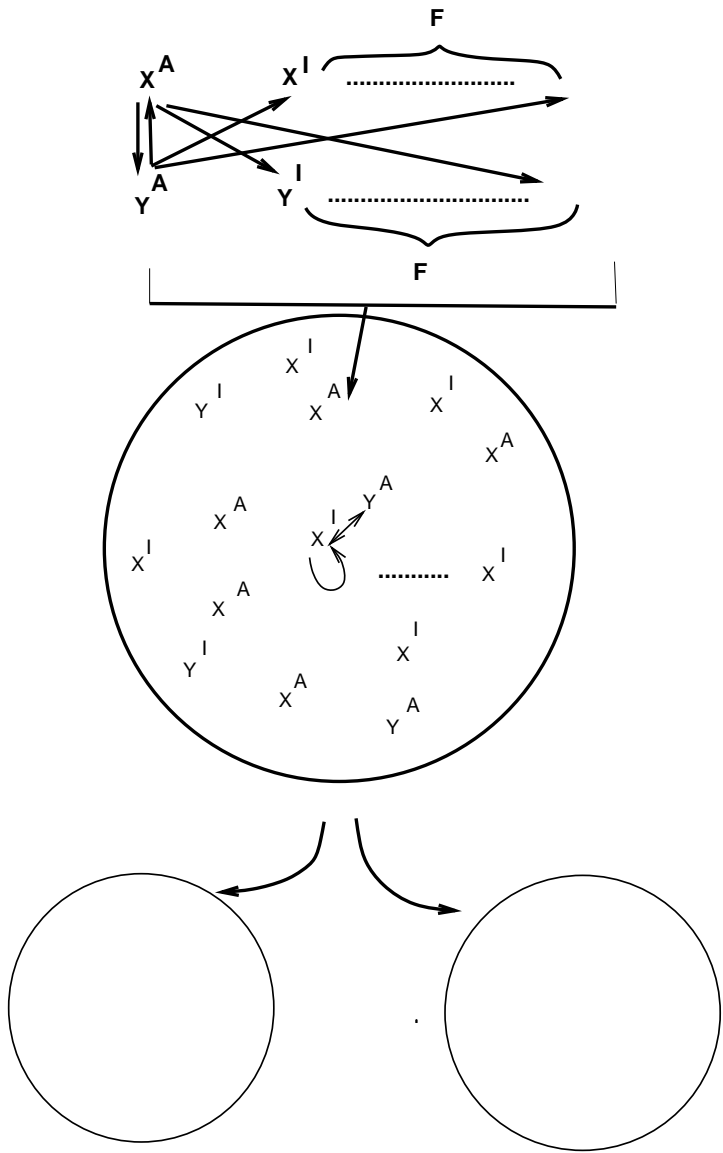,width=.7\textwidth}
\caption{Schematic representation of our model}
\end{figure}

\begin{figure}
\noindent
\hspace{-.3in}
\epsfig{file=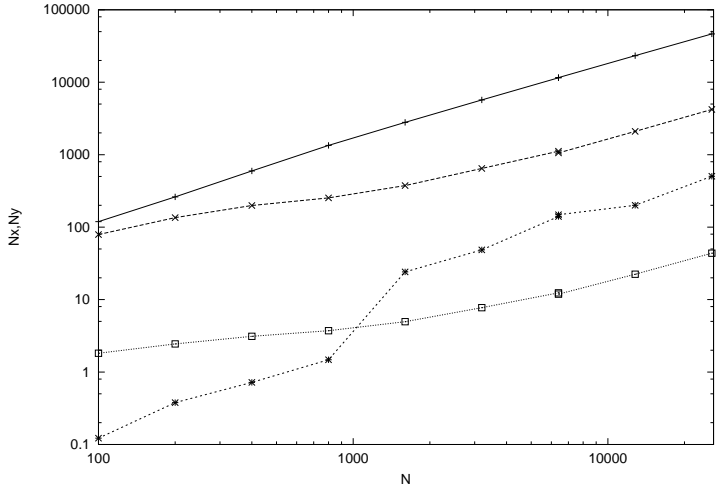,width=.7\textwidth,angle=-90}
\caption{
Dependence of $\langle N_x^A \rangle (\times)$, $\langle N_x^I \rangle (+)$, 
$\langle N_y^A \rangle (\Box)$, and $\langle N_y^I \rangle(*)$ 
on $N$.
The parameters were fixed as $\gamma_x=1$, $\gamma_y=0.01$, and $\mu =.05$.
Plotted are the averages of $N_x^A$, $N_x^I$, $N_y^A$, and $N_y^I$
at the division event, and thus their sum is
$2N$.  
In all the simulations conducted for the present paper, 
we used $M_{tot}=100$, and 
the sampling for the averages were taken over $10^5-3\times 10^5$ steps,
where the number of divisions ranges from $10^4$ to $10^5$,
depending on the parameters.}
\end{figure}

\begin{figure}
\noindent
\hspace{-.3in}
\epsfig{file=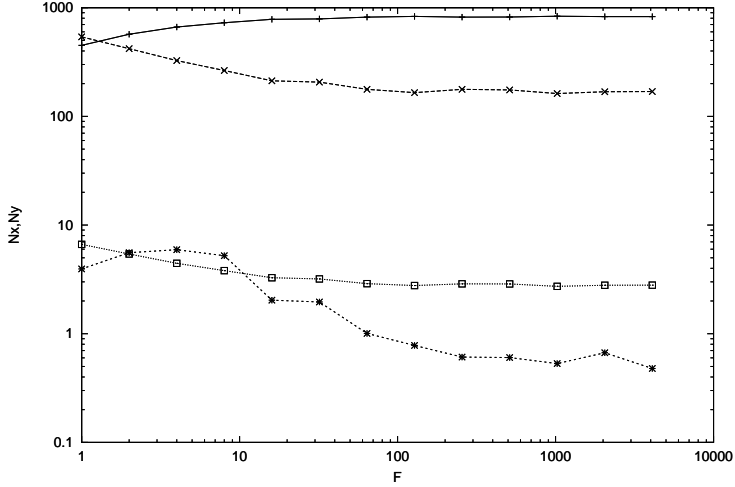,width=.7\textwidth,angle=-90}
\caption{
Dependence of $\langle N_x^A \rangle (\times)$, $\langle N_x^I \rangle (+)$, 
$\langle N_y^A \rangle (\Box)$, and $\langle N_y^I \rangle(*)$ 
on $F$.
The parameters were fixed as 
$\gamma_x=1$, $\gamma_y=.01$, $\mu =.05$, and $N=1000$.
Plotted are the averages of $N_x^A$, $N_x^I$, $N_y^A$, and $N_y^I$
at the division event, and thus their sum is
$2N=2000$.}
\end{figure}

\begin{figure}
\noindent
\hspace{-.3in}
\epsfig{file=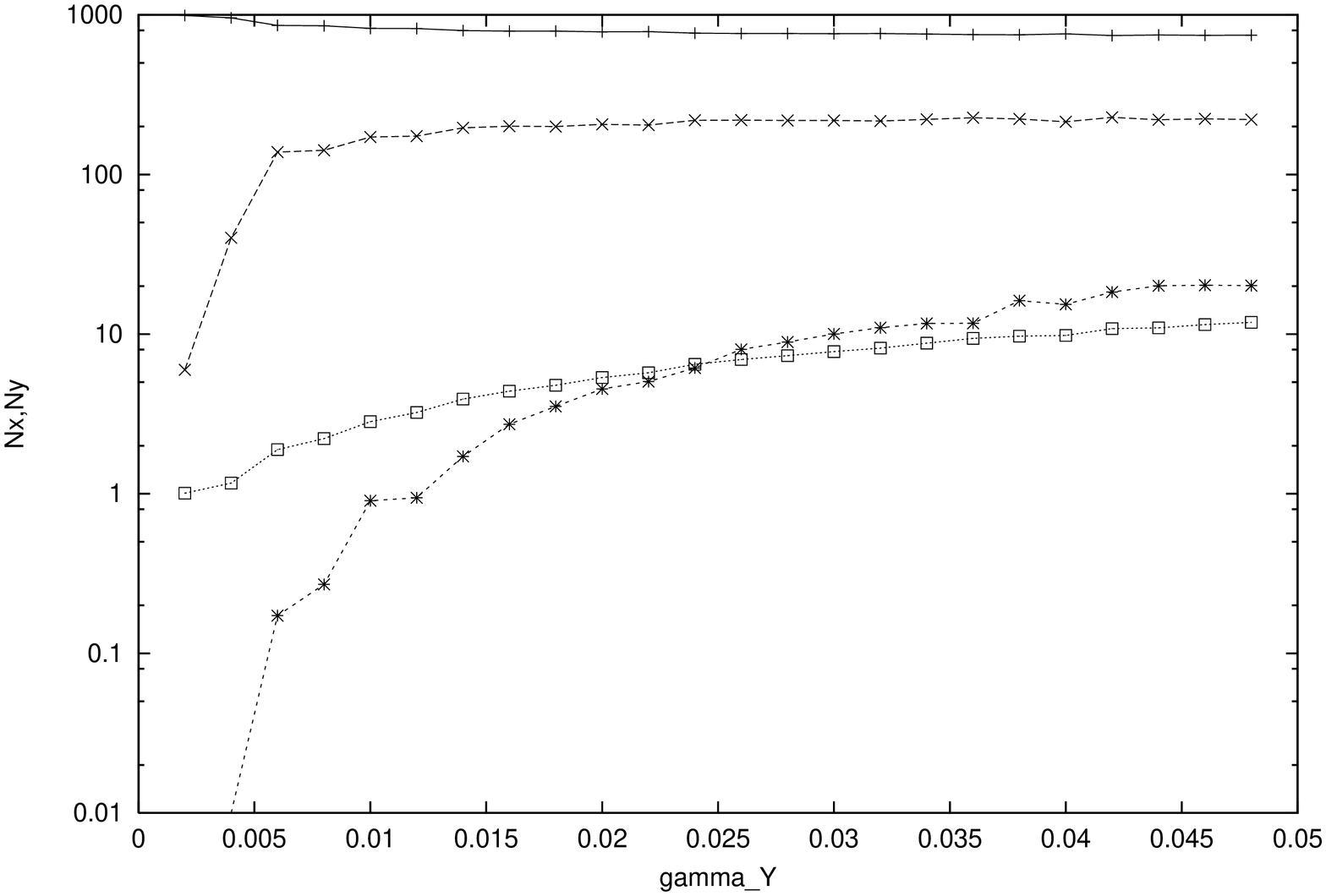,width=.7\textwidth,angle=-90}
\caption{
Dependence of $\langle N_x^A \rangle (\times)$, $\langle N_x^I \rangle (+)$, 
$\langle N_y^A \rangle (\Box)$, and $\langle N_y^I \rangle(*)$ 
on $\gamma_y$.
The parameters were fixed as 
$\gamma_x=1$, $\mu =.05$, $F=128$, and $N=1000$.
Plotted are the averages of $N_x^A$, $N_x^I$, $N_y^A$, and $N_y^I$
at the division event.}
\end{figure}

\begin{figure}
\noindent
\hspace{-.3in}
\epsfig{file=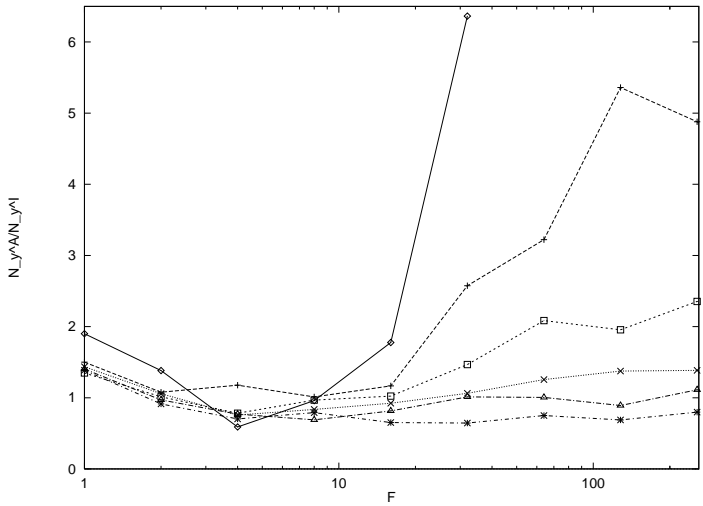,width=.7\textwidth,angle=-90}
\caption{
Dependence of the active-to-inactive ratio, $\frac{\langle N_y^A \rangle }{\langle N_y^I \rangle }$,
on $F$.
The parameters were fixed as $\gamma_x=1$, $\gamma_y=.01$, $\mu =.05$, and $F=128$.
Plots for $\gamma_y=.005$ ($\Diamond$), .01 (+), .015 ($\Box$), 0.02 ($\times$), 0.025 ($\triangle$),
and 0.03 (*) are overlaid.  
Plotted are the averages of $N_x^A$, $N_x^I$, $N_y^A$, and $N_y^I$
at the division event.}
\end{figure}


\begin{figure}
\noindent
\hspace{-.3in}
\epsfig{file=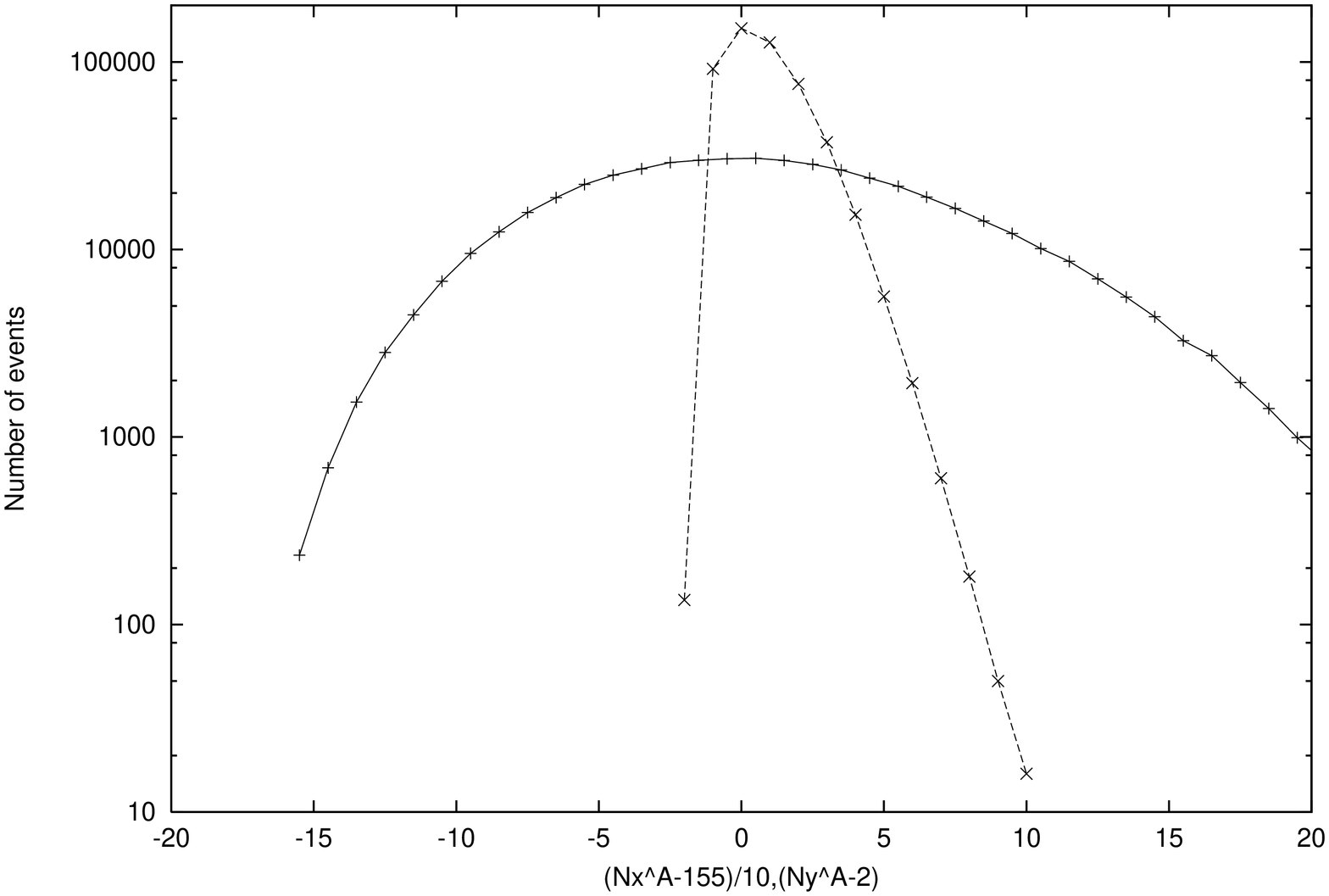,width=.6\textwidth,angle=-90}
\epsfig{file=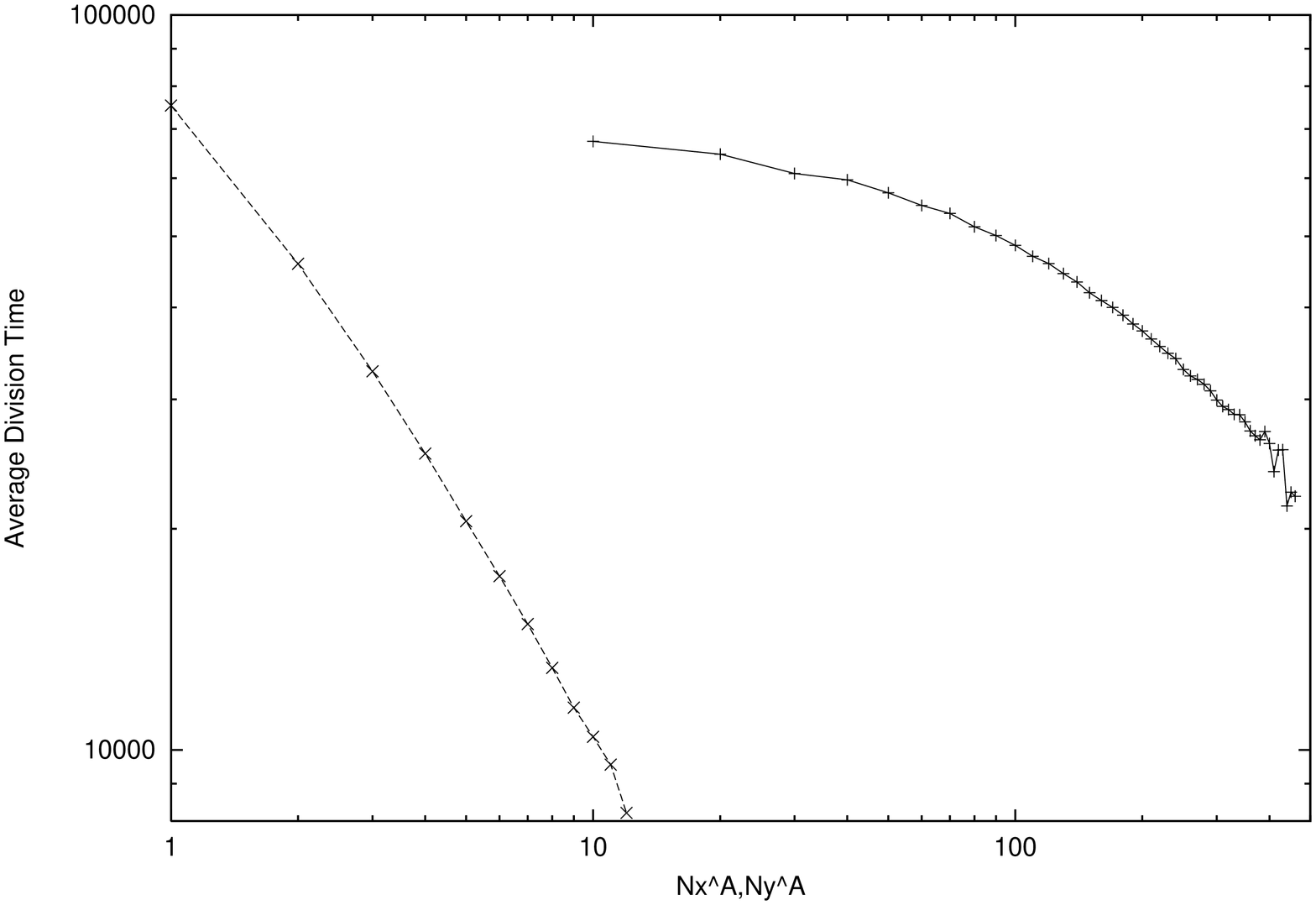,width=.6\textwidth,angle=-90}
\caption{
(a) Histograms of the number of division events with given $N_x$ (+) and with
given $N_y$ ($\times$), 
plotted versus $\frac{N^A_x-155}{10}$ and $N^A_y-2$, respectively. 
The histogram representing the averages for $N_x$ was
computed with a bin size of 10 as discussed in the text.
These histograms were found with a sampling of  
divisions occurring between the  $6\times 10^5$ and $10^6$ time steps.
(b) Average number of time steps $\overline{T_d}$ required for the division of protocells for given
$N_y$($\times$) and $N_x$ (+). These plots were obtained  using the histograms in Fig.6 (a).
}
\end{figure}

\begin{figure}
\noindent
\hspace{-.3in}
\epsfig{file=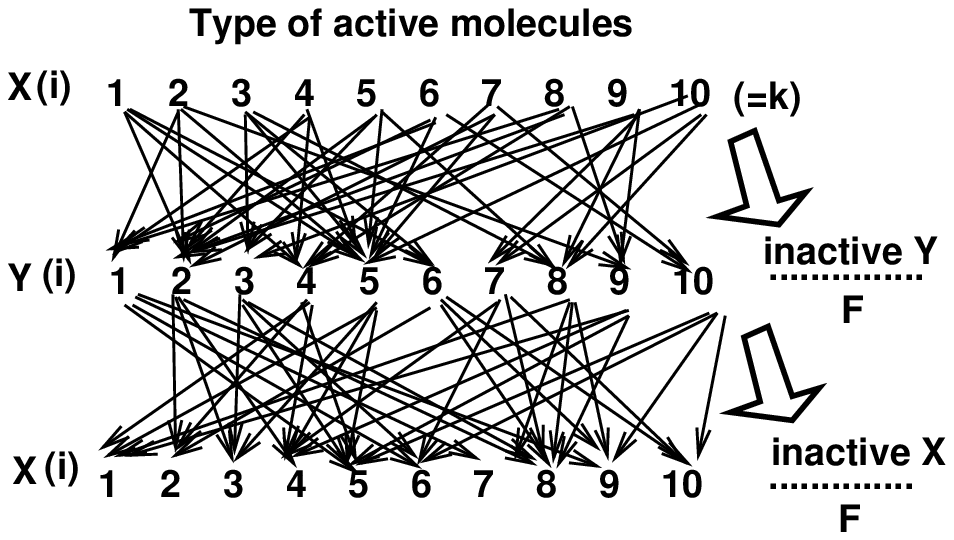,width=.6\textwidth}
\epsfig{file=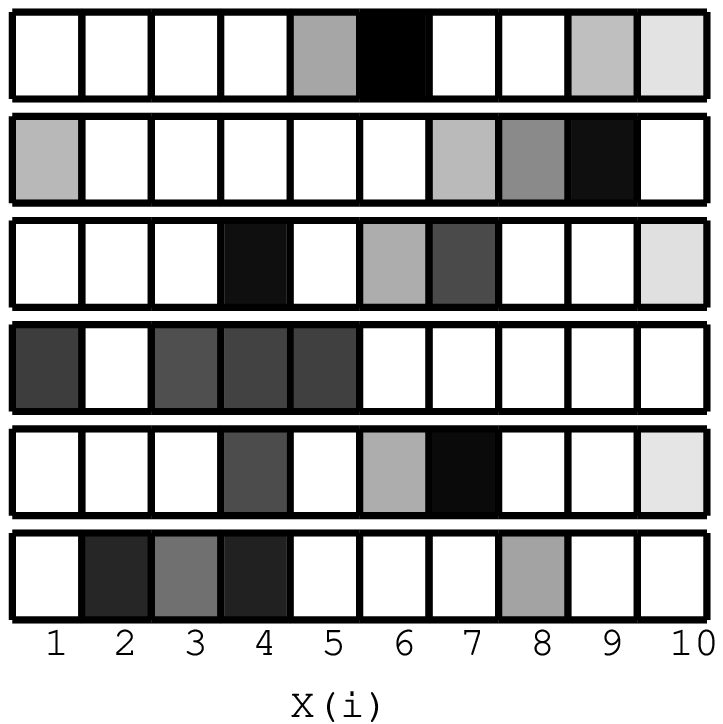,width=.2\textwidth}
\caption{(a) Catalytic network between $X(i)$ and $Y(i)$.  
The arrows from the top column points to types of the 
$Y$ species that each $X(i)$ catalyzes, while those from the middle
points to types of the $X$ species that each $Y(i)$ catalyzes.
Each of the $k=10$ active species catalyzes the synthesis of 4 chemicals.
(b) The logarithm of the average population of $X(i)$ displayed wish a
gray scale.
From top to bottom, 6 samples resulting from different initial conditions are plotted.
The type $i_r$ of $Y(i)$ with non-vanishing population 
$Y(i)$ corresponding to each column is as follows (top to bottom)
$i_r=10$, 8, 4, 5, 4, and 2.
The parameters were chosen as $F=16\times k=160$, $\mu =0.03$, $\gamma_x=1$, 
$\gamma_y=0.01$, $N=1000$.}
\end{figure}

\begin{figure}
\noindent
\hspace{-.3in}
\epsfig{file=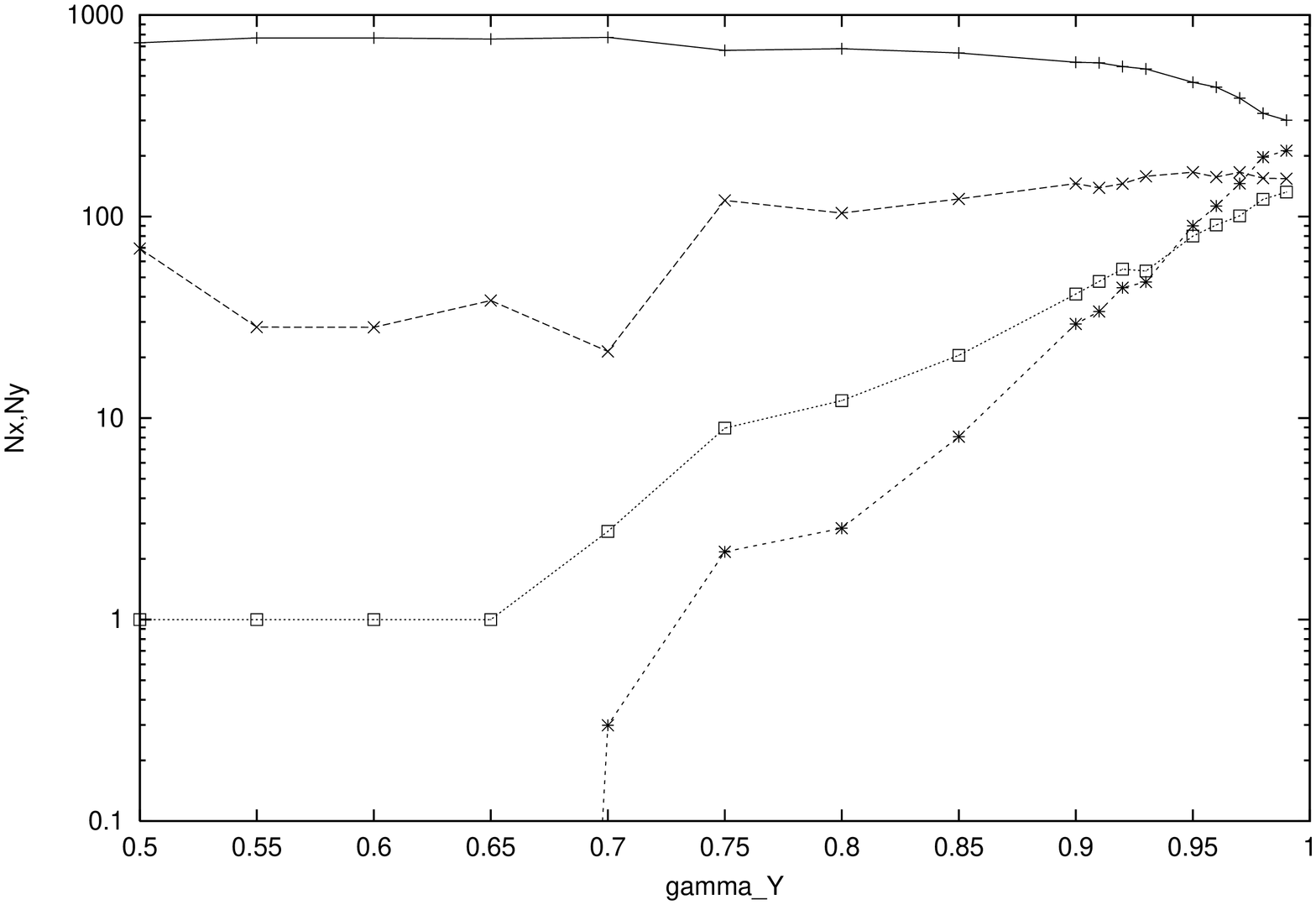,width=.7\textwidth,angle=-90}
\caption{
Dependence of $\langle N_x^A \rangle (\times)$, $\langle N_x^I \rangle (+)$, 
$\langle N_y^A \rangle (\Box)$, and $\langle N_y^I \rangle(*)$ 
on $\gamma_y$.
The parameters were fixed as 
$\gamma_x=1$, $\mu =.05$, $F=128$, and $N=500$.
Plotted are the averages of $N_x^A$, $N_x^I$, $N_y^A$, and $N_y^I$
at the division event.}
\end{figure}

\end{document}